\documentclass[conference]{IEEEtran}
\IEEEoverridecommandlockouts
\usepackage{cite}
\usepackage{amsmath,amssymb,amsfonts}
\usepackage{algorithmic}
\usepackage{array}
\usepackage{graphicx}
\usepackage{textcomp}
\usepackage{xcolor}
\usepackage[hidelinks]{hyperref}
\usepackage[all]{hypcap}
\usepackage{balance}
\newcommand{\figref}[1]{\hyperref[#1]{Fig.~\ref*{#1}}}
\newcommand{\tabref}[1]{\hyperref[#1]{Table~\ref*{#1}}}
\newcommand{\eqnref}[1]{\hyperref[#1]{\textup{(\ref*{#1})}}}

\def\BibTeX{{\rm B\kern-.05em{\sc i\kern-.025em b}\kern-.08em
		T\kern-.1667em\lower.7ex\hbox{E}\kern-.125emX}}

\begin{document}
	
	\title{\LARGE Fluid-Antenna-Assisted Distributed Joint Decoding for \\Cell-Free Massive MIMO Unsourced Random Access}
	
	\author{Liandong Hu, Jian Dang, Zaichen Zhang
		\thanks{This work is supported by the Fundamental Research Funds for the Central Universities (2242022k60001) ,  Basic Research Program of Jiangsu (No. BK20252003)}
		
		\thanks{Liandong Hu is with the National Mobile Communications Research Laboratory, Frontiers Science Center for Mobile Information Communication and Security,	Southeast University, Nanjing, 211189, China. (e-mail: liandonghu@seu.edu.cn).}
		
		\thanks{Jian Dang is with the National Mobile Communications Research Labo-ratory, Frontiers Science Center for Mobile Information Communication and Security, Southeast University, Nanjing 211189, China and he is also with Purple Mountain Laboratories, Nanjing 211111, China.(email: dangjian@seu.edu.cn).  \emph{Corresponding Author: Jian Dang.}}
		
		\thanks{Zaichen Zhang is with the National Mobile Communications Research Laboratory,Frontiers Science Center for Mobile Information Communication and Security, Southeast University, Nanjing, 210096, China and he is also	with the Purple Mountain Laboratories, Nanjing 211111, China (e-mail:	zczhang@seu.edu.cn). }

	}
	
	\maketitle
	
	\begin{abstract}
		Unsourced random access (URA) supports massive sporadic connectivity but remains limited by multiuser interference, distributed channel uncertainty, and local fading. This paper proposes a fluid-antenna-assisted distributed joint decoder for cell-free massive MIMO URA. Single-radio-frequency-chain access points sequentially sound correlated ports; the central processor then performs correlation-aware expectation-maximization approximate message passing, selects one common payload port per access point, and combines MIMO iterative Gaussian approximation with full-packet channel re-estimation and successive interference cancellation. Fixed-port ablations confirm the complementary benefits of distributed reception, re-estimation, and iterative cancellation, while fluid-antenna simulations show improved average and lower-tail received power over fixed-port reception. The framework provides a feasible integration of fluid antennas and cell-free URA and motivates end-to-end evaluation under pilot collisions and switching overhead.
	\end{abstract}
	
	
	\section{INTRODUCTION}
	
	\subsection{Background and Related Work}
	
	URA removes scheduling and device-identity overhead by letting all active terminals share a codebook and asking the receiver for an unordered message list \cite{polyanskiy2017perspective}. Coded compressed sensing couples sparse recovery with channel decoding \cite{amalladinne2022ccs}, while random-spreading URA uses a message-selected sequence and iterative soft interference cancellation. Multiple antennas improve separation; distributing them as cell-free access points (APs) further supplies macrodiversity and reduces cell-edge effects \cite{ngo2017cellfree}, although fixed AP antennas remain vulnerable to local fading \cite{zhang2026codedpattern,zhang2026lessismore,zhang2024randomscattering,zhang2024odma,zhang2026isacura,zhang2025probabilisticodma,zhang2025geographical,zhang2025uncoupled}.
	
	Fluid antenna systems (FASs) connect one radio-frequency (RF) chain to a selectable port within a compact aperture, thereby exploiting favorable spatial samples without one chain per port \cite{wong2021fas,wong2022fama}. Closely spaced ports are correlated, so independent-port models overstate diversity. Recent AP-side cell-free FAS and FAS-assisted URA studies \cite{olyaee2024apfas,zhang2025fasura} motivate combining port reconfiguration with sparse coded detection; a correlation-aware receiver for random-spreading cell-free URA is still needed \cite{meng2025cooperativeanalysis,wu2026scalablefas,wu2026blockcorrelation,chen2024beamtraining,xu2025rydberg,xu2024nearfield,zhang2026isaclimits,zhang2026finiteblocklength,zhang2026activityangular,zhang2026finiteblockcorrelation,zhang2026finiteaperture,zhang2026crb,zhang2026dualside,zhang2026jointpattern}.
	
	\subsection{Challenges and Contributions}
	
	The CPU must jointly infer sparse activity and correlated port channels, then choose one physically common port per single-RF-chain AP before payload reception. The selected ports remain fixed during decoding and successive interference cancellation (SIC), while activity errors, code correlations, near--far effects, and channel uncertainty interact.
	
	The main contributions of this paper are summarized as follows.
	\begin{itemize}
		\item A single-RF-chain FAS-assisted cell-free URA model captures sequential sounding and spatial correlation across $Q$ candidate ports.
		\item Correlation-aware EM-AMP jointly detects activity and estimates port channels; posterior aggregate energy selects a common AP port for MIMO-IGA/NR-LDPC Turbo decoding.
		\item Full-packet regularized least-squares (RLS) re-estimation and SIC cancel decoded pilots at all sounded ports and payloads at the fixed selected ports.
	\end{itemize}
	
	The remainder of the paper is organized as follows: Section~II defines the system model. The proposed fluid-antenna-assisted distributed joint decoder is introduced in Section~III. Numerical results are presented in Section~IV, and Section~V concludes the paper.
	
	\textit{Notation:} Italic, bold lowercase, bold uppercase, and calligraphic letters denote scalars, vectors, matrices, and sets. Hats and overbars denote estimates and selected-port quantities. $(\cdot)^{\mathrm T}$, $(\cdot)^{\mathrm H}$, $(\cdot)^*$, $\|\cdot\|_{\mathrm F}$, $\operatorname{tr}(\cdot)$, $\operatorname{blkdiag}(\cdot)$, $\operatorname{vec}(\cdot)$, and $\otimes$ denote transpose, Hermitian transpose, conjugation, Frobenius norm, trace, block diagonalization, vectorization, and Kronecker product. $[\mathbf X]_{i,j}$ denotes the $(i,j)$th entry; $|\cdot|$ denotes magnitude/cardinality. $\mathbf0$, $\mathbf I$, $\mathbb C$, and $\mathbb R$ denote zero, identity, complex, and real domains. $\mathbb E[\cdot]$, $\operatorname{Var}(\cdot)$, $\Pr(\cdot)$, and $\operatorname{Re}\{\cdot\}$ denote expectation, variance, probability, and real part. $\mathcal{CN}(\mathbf x;\boldsymbol\mu,\mathbf C)$ and $\delta(\cdot)$ denote the proper complex Gaussian PDF and Dirac delta function.

	\section{SYSTEM MODEL}
	\label{sec:system_model}
	
	Consider $M$ distributed APs serving $K_a$ active single-antenna UEs in a disk of radius $R$. Each AP connects one RF chain to $Q$ switchable ports on a linear aperture of length $W\lambda_c$, where $\lambda_c$ is the carrier wavelength. The all-port channel vector from UE $u$ to AP $m$ is
	
	\begin{equation}
		\label{eq:fas_channel_vector}
		\mathbf{h}_{u,m}
		=[h_{u,m,1},\ldots,h_{u,m,Q}]^{\mathrm{T}}
		=\sqrt{\phi_{u,m}}\mathbf{g}_{u,m}
	\end{equation}
	with $\phi_{u,m}=(d_{u,m}/d_0)^{-\beta}$ constant across the compact aperture. Under two-dimensional isotropic scattering,
	\begin{equation}
		\label{eq:fas_correlation}
		\mathbf{g}_{u,m}\sim\mathcal{CN}(\mathbf{0},\mathbf{R}_{Q}),~
		[\mathbf{R}_{Q}]_{q,r}
		=J_0\!\left(\frac{2\pi W|q-r|}{Q-1}\right),
	\end{equation}
	where $J_0(\cdot)$ is the zeroth-order Bessel function. UE--AP pairs are independent, port channels within an AP are correlated, and fading is block-flat. For $Q=1$, $\mathbf R_Q=[1]$.
	
	Each $B_{\mathrm{info}}$-bit message uses its first $B_p$ bits to select a unit-norm column of the common Gaussian codebook $\mathbf A\in\mathbb C^{n_p\times K}$, $K=2^{B_p}$. The remaining bits are protected by CRC and NR-LDPC coding \cite{3gpp38212} and BPSK-modulated to $\mathbf s_u\in\{+1,-1\}^{n_c}$. Active indices $\mathcal D$ are assumed distinct; collision resolution is outside the present scope.
	
	The frame has $Q$ sequential port-sounding pilot subslots and $n_c$ payload slots. In pilot subslot $q$, AP $m$ observes
	\begin{equation}
		\mathbf y_{m,q,0}=\sum_{u=1}^{K_a}h_{u,m,q}\mathbf a_{i_u}
		+\mathbf n_{m,q,0},~ q=1,\ldots,Q.
		\label{eq:fas_received_pilot}
	\end{equation}
	Stacking all AP-port observations gives the all-port row-sparse model
	\begin{equation}
		\label{eq:fas_pilot_mmv}
		\mathbf{Y}_0
		=\mathbf{A}\mathbf{H}+\mathbf{N}_0,
	\end{equation}
	where the all-port channel matrix $\mathbf H\in\mathbb C^{K\times MQ}$ has nonzero rows only for $i\in\mathcal D$, and noise entries have variance $\sigma_n^2$.
	Equivalently,
	\begin{equation}
		[\mathbf H]_{i,(m-1)Q+q}=
		\begin{cases}
			h_{u,m,q}, & i=i_u\text{ for some }u,\\
			0, & i\notin\mathcal D.
		\end{cases}
		\label{eq:fas_channel_entries}
	\end{equation}
	
	After sounding, the CPU selects one port $q_m^\star$ per AP and holds it fixed for all payload slots. With $\bar h_{u,m}=h_{u,m,q_m^\star}$, AP $m$ receives
	\begin{equation}
		\mathbf y_{m,j}=\sum_{u=1}^{K_a}\bar h_{u,m}\mathbf a_{i_u}s_{u,j}
		+\mathbf n_{m,j},~ j=1,\ldots,n_c,
		\label{eq:fas_received_data}
	\end{equation}
	and the concatenated observation is
	\begin{equation}
		\label{eq:cf_received_matrix}
		\mathbf{Y}_j=\mathbf{X}_j\overline{\mathbf{H}}+\mathbf{N}_j,
	\end{equation}
	where $\mathbf X_j=[\mathbf a_{i_1}s_{1,j},\ldots,\mathbf a_{i_{K_a}}s_{K_a,j}]$ and $\overline{\mathbf H}\in\mathbb C^{K_a\times M}$ contains selected-port channels.
	
	One access attempt occupies $L=n_p(Q+n_c)$ channel uses. Under unit-energy spreading and BPSK symbols, the energy-per-source-bit convention is $E_b/N_0=(Q+n_c)/(B_{\mathrm{info}}\sigma_n^2)$. Without identifying the active UEs, the receiver generates an estimated list $\widetilde{\mathcal L}$ from $[\mathbf Y_0,\mathbf Y_1,\ldots,\mathbf Y_{n_c}]$. The URA performance metric is the per-user probability of error (PUPE), $P_e=P_{\mathrm{md}}+P_{\mathrm{fa}}$, where
	\begin{equation}
		P_{\mathrm{md}}=\frac{1}{K_a}
		\sum_{\mathbf u_u\in\mathcal L}
		\Pr(\mathbf u_u\notin\widetilde{\mathcal L}),~
		P_{\mathrm{fa}}=
		\frac{|\widetilde{\mathcal L}\setminus\mathcal L|}
		{|\widetilde{\mathcal L}|}.
		\label{eq:cf_pupe_metrics}
	\end{equation}
	
	\section{DISTRIBUTED JOINT DECODING SCHEME}
	\label{sec:joint_decoding}
	
	The proposed receiver has four stages. First, the CPU uses all $MQ$ port-sounding observations for joint activity detection and correlated port-channel estimation. Second, it selects one common receive port at each AP and fixes that port for the data phase. Third, a MIMO-IGA Turbo decoder exchanges soft information with an NR-LDPC decoder over the resulting $M$-dimensional effective channel \cite{hu2026iga}. Finally, successfully decoded packets undergo channel re-estimation and SIC, after which the residual pilot observations are processed again.
	
	\subsection{Correlation-Aware EM-AMP Front End}
	\label{subsec:em_amp}
	
	Let $P=MQ$ and let $\mathbf{h}_i\in\mathbb{C}^{P}$ be the transpose of row $i$ of the all-port matrix $\mathbf{H}$. The indicator $\alpha_i$ equals one when codeword $i$ is occupied and zero otherwise. Building on generalized AMP and its EM extensions \cite{vila2013emamp}, we use a vector denoiser because all entries of an occupied row share the same support event and the $Q$ ports of each AP are spatially correlated \cite{zhu2023temporally}:
	\begin{equation}
		\label{eq:fas_bg_prior}
		p(\mathbf{h}_i)
		=(1-\lambda_i)\delta(\mathbf{h}_i)
		+\lambda_i\mathcal{CN}(\mathbf{h}_i;
		\mathbf{0},\mathbf{C}_i),
	\end{equation}
	where $\lambda_i=\Pr(\alpha_i=1)$. Let
	$\mathbf R_{Q,\epsilon}=\mathbf R_Q+\epsilon\mathbf I$
	with $\epsilon>0$. The regularized active-row covariance is
	\begin{equation}
		\label{eq:fas_prior_covariance}
		\mathbf{C}_i=\operatorname{blkdiag}
		(\phi_{i,1}\mathbf{R}_{Q,\epsilon},\ldots,
		\phi_{i,M}\mathbf{R}_{Q,\epsilon}).
	\end{equation}
	The scales $\phi_{i,m}$ are learned by EM because codeword hypotheses have no user identities. With $p=(m-1)Q+q$, the complex GAMP linear step is compactly written as
	\begin{align}
		\mu_{l,p}^{p}&=\sum_i|[\mathbf A]_{l,i}|^2[\widehat{\mathbf V}_i]_{p,p},\notag\\
		\hat p_{l,p}&=\sum_i[\mathbf A]_{l,i}\hat h_{i,p}-\mu_{l,p}^{p}\hat u_{l,p}^{-},\notag\\
		\hat u_{l,p}&=\frac{[\mathbf Y_0]_{l,p}-\hat p_{l,p}}
		{\mu_{l,p}^{p}+\sigma_n^2},\notag\\
		\mu_{i,p}^{r}&=\left(\sum_l\frac{|[\mathbf A]_{l,i}|^2}
		{\mu_{l,p}^{p}+\sigma_n^2}\right)^{-1},\notag\\
		\hat r_{i,p}&=\hat h_{i,p}+\mu_{i,p}^{r}\sum_l[\mathbf A]_{l,i}^*\hat u_{l,p}.
		\label{eq:fas_gamp_linear}
	\end{align}
	These updates yield $\mathbf r_i^{(t)}=\mathbf h_i+\mathbf v_i^{(t)}$, where $\mathbf v_i^{(t)}\sim\mathcal{CN}(\mathbf0,\boldsymbol\Sigma_i^{r,(t)})$ and $\boldsymbol\Sigma_i^{r,(t)}$ is diagonal. The correlation-aware vector denoiser evaluates
	\begin{equation}
		\label{eq:fas_posterior_activity}
		\pi_i^{(t)}=\left[1+\frac{1-\lambda_i^{(t)}}{\lambda_i^{(t)}}
		\frac{\mathcal{CN}(\mathbf{r}_i^{(t)};\mathbf{0},
			\boldsymbol{\Sigma}_i^{r,(t)})}
		{\mathcal{CN}(\mathbf{r}_i^{(t)};\mathbf{0},
			\mathbf{C}_i^{(t)}+\boldsymbol{\Sigma}_i^{r,(t)})}
		\right]^{-1}.
	\end{equation}
	Conditioned on activity, the posterior mean and covariance are
	\begin{align}
		\boldsymbol{\gamma}_i^{(t)}
		&=\mathbf{C}_i^{(t)}(\mathbf{C}_i^{(t)}+
		\boldsymbol{\Sigma}_i^{r,(t)})^{-1}\mathbf{r}_i^{(t)},
		\label{eq:fas_conditional_mean}\\
		\mathbf{V}_i^{(t)}
		&=\mathbf C_i^{(t)}-\mathbf C_i^{(t)}
		(\mathbf C_i^{(t)}+\boldsymbol{\Sigma}_i^{r,(t)})^{-1}
		\mathbf C_i^{(t)}.
		\label{eq:fas_conditional_covariance}
	\end{align}
	The unconditional AMP estimates are
	\begin{align}
		\widehat{\mathbf{h}}_i^{(t+1)}
		&=\pi_i^{(t)}\boldsymbol{\gamma}_i^{(t)},
		\label{eq:fas_posterior_mean}\\
		\widehat{\mathbf{V}}_i^{(t+1)}
		&=\pi_i^{(t)}(\mathbf{V}_i^{(t)}+
		\boldsymbol{\gamma}_i^{(t)}\boldsymbol{\gamma}_i^{(t)\mathrm{H}})
		-\widehat{\mathbf{h}}_i^{(t+1)}
		\widehat{\mathbf{h}}_i^{(t+1)\mathrm{H}}.
		\label{eq:fas_posterior_covariance}
	\end{align}
	The EM stage sets $\lambda_i^{(t+1)}=\pi_i^{(t)}$ and updates each AP-dependent scale as
	\begin{equation}
		\label{eq:fas_em_scale}
		\phi_{i,m}^{(t+1)}=\frac{1}{Q}\operatorname{tr}\!\left[
		\mathbf{R}_{Q,\epsilon}^{-1}
		(\mathbf{V}_{i,m}^{(t)}+
		\boldsymbol{\gamma}_{i,m}^{(t)}
		\boldsymbol{\gamma}_{i,m}^{(t)\mathrm{H}})\right],
	\end{equation}
	where the subscript $m$ selects the $Q\times Q$ AP block. \hyperref[alg:em_amp]{Algorithm~1} summarizes the alternating updates.
	
	\capstartfalse
	\begin{figure}[t!]
		\phantomsection\label{alg:em_amp}
		\noindent\begin{minipage}{\columnwidth}
			\hrule height 1pt
			\vspace{0.06cm}
			\noindent\textbf{Algorithm 1: Correlation-Aware EM-AMP for FAS-Assisted Cell-Free Access}
			\vspace{0.04cm}
			\hrule height 0.5pt
			\vspace{0.04cm}
			\scriptsize
			\renewcommand{\arraystretch}{1.04}
			\noindent\textbf{Input:} $\mathbf{Y}_0,\mathbf{A},K,M,Q,
			n_p,\sigma_n^2,\lambda_{\mathrm{th}},T_{\max},\tau_{\mathrm{th}},
			\epsilon$.
			\vspace{0.03cm}
			
			\begin{tabular}{@{}r@{\hspace{0.3em}}>{\raggedright\arraybackslash}p{0.88\columnwidth}@{}}
				\textbf{1} & Initialize $\lambda_i$, $\phi_{i,m}$,
				$\widehat{\mathbf h}_i=\mathbf0$, and $\widehat{\mathbf V}_i$ from pilot energy.\\
				\textbf{2} & \textbf{for} $t=1,\ldots,T_{\max}$ \textbf{do}\\
				\textbf{3} & \quad Apply complex GAMP to form $\mathbf r_i^{(t)}$ and
				$\boldsymbol{\Sigma}_i^{r,(t)}$.\\
				\textbf{4} & \quad Evaluate
				$\pi_i^{(t)}$, $\boldsymbol{\gamma}_i^{(t)}$, and
				$\mathbf V_i^{(t)}$ from
				\eqnref{eq:fas_posterior_activity}--\eqnref{eq:fas_conditional_covariance}.\\
				\textbf{5} & \quad Update $\widehat{\mathbf h}_i^{(t+1)}$ and
				$\widehat{\mathbf V}_i^{(t+1)}$ using
				\eqnref{eq:fas_posterior_mean} and
				\eqnref{eq:fas_posterior_covariance}.\\
				\textbf{6} & \quad EM: set $\lambda_i^{(t+1)}=\pi_i^{(t)}$
				and update $\phi_{i,m}^{(t+1)}$ using
				\eqnref{eq:fas_em_scale}.\\
				\textbf{7} & \quad Stop if
				$\tau(t+1)\leq\tau_{\mathrm{th}}$.\\
				\textbf{8} & \textbf{end for}\\
				\textbf{9} & \textbf{Output:}
				$\widehat{\mathcal K}=\{i:\lambda_i\geq\lambda_{\mathrm{th}}\}$,
				$\widehat{\mathbf H}$, and posterior statistics.\\
			\end{tabular}
			\vspace{0.04cm}
			\hrule height 1pt
		\end{minipage}
	\end{figure}
	\capstarttrue
	
	Iterations stop at $T_{\max}$ or when $\tau=\|\widehat{\mathbf H}^{(t+1)}-\widehat{\mathbf H}^{(t)}\|_{\mathrm F}^2/\max\{\|\widehat{\mathbf H}^{(t)}\|_{\mathrm F}^2,\epsilon\}$ falls below $\tau_{\mathrm{th}}$. A codeword is active when its final posterior probability exceeds $\lambda_{\mathrm{th}}$.
	
	\subsection{Fluid-Antenna Port Selection}
	\label{subsec:port_selection}
	
	For the final posterior blocks $\boldsymbol\gamma_{i,m}$ and $\mathbf V_{i,m}$, AP $m$ selects one common port and extracts each detected effective channel as
	\begin{align}
		\eta_{m,q}&=\sum_{i=1}^{K}\pi_i
		(|\gamma_{i,m,q}|^2+[\mathbf V_{i,m}]_{q,q}),
		\label{eq:fas_port_score}\\
		q_m^\star&=\arg\max_q\eta_{m,q},~
		\hat{\bar h}_{i,m}=[\widehat{\mathbf h}_i]_{(m-1)Q+q_m^\star}.
	\end{align}
	The selected-port rows form $\widehat{\overline{\mathbf H}}$. Port indices remain fixed during payload decoding and SIC; only their channel coefficients are refined.
	
	\subsection{MIMO-IGA-Based Turbo Decoder}
	\label{subsec:mimo_iga}
	
	After port selection, the CPU iterates an elementary signal estimator (ESE) with the NR-LDPC decoder over support $\widehat{\mathcal K}$. Joint code matching and spatial MRC reduce \eqnref{eq:cf_received_matrix} to
	\begin{align}
		z_{k,j}&=\mu_kx_{k,j}+\sum_{i\ne k}\rho_{k,i}x_{i,j}+\widetilde n_{k,j},
		\label{eq:mimo_compact_observation}\\
		\mu_k&=\overline{\mathbf h}_k^{\mathrm H}\overline{\mathbf h}_k,~
		\rho_{k,i}=(\mathbf a_k^{\mathrm H}\mathbf a_i)
		(\overline{\mathbf h}_k^{\mathrm H}\overline{\mathbf h}_i),
		\notag
	\end{align}
	where $\mathbb E|\widetilde n_{k,j}|^2=\mu_k\sigma_n^2$. Approximating residual interference as complex Gaussian gives
	\begin{align}
		m_{k,j}^{I}&=\sum_{i\ne k}\rho_{k,i}\mathbb E[x_{i,j}],~
		v_{k,j}^{I}=\sum_{i\ne k}|\rho_{k,i}|^2\operatorname{Var}(x_{i,j})
		+\mu_k\sigma_n^2. \label{eq:mimo_interf_mean}
	\end{align}
	Thus $z_{k,j}|x_{k,j}\sim\mathcal{CN}(\mu_kx_{k,j}+m_{k,j}^{I},v_{k,j}^{I})$, and the resulting ESE extrinsic LLR is
	\begin{equation}
		\label{eq:mimo_llr_simplified}
		L_{k,j}^{\mathrm{ESE}}
		=\frac{4\mu_k\operatorname{Re}\{z_{k,j}-m_{k,j}^{I}\}}{v_{k,j}^{I}}.
	\end{equation}
	The decoder returns extrinsic information and the next ESE iteration uses
	\begin{align}
		\label{eq:mimo_ext_info}
		L_{k,j}^{\mathrm{Ext}}&=L_{k,j}^{\mathrm{DEC}}-L_{k,j}^{\mathrm{ESE}},\\
		\mathbb E[x_{k,j}]&=\tanh(L_{k,j}^{\mathrm{Ext}}/2),~
		\operatorname{Var}(x_{k,j})=1-\mathbb E[x_{k,j}]^2.\notag
	\end{align}
	
	\begin{figure*}[!t]
		\centering
		\includegraphics[width=0.8\textwidth]{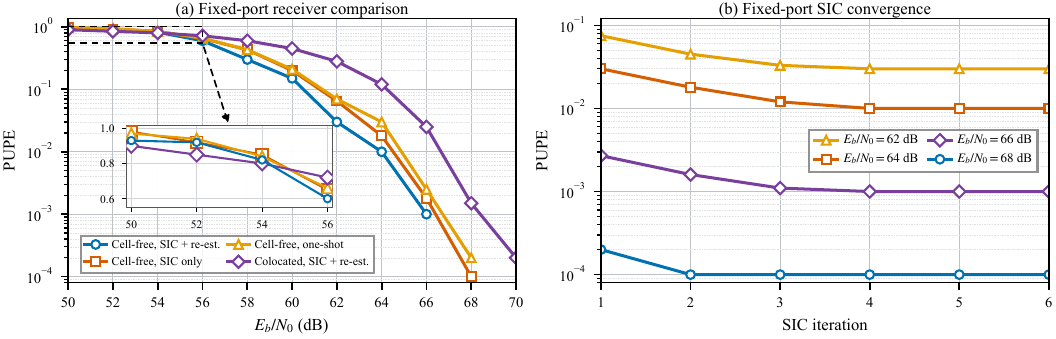}
		\caption{Archived fixed-port ($Q=1$) results for $K_a=100$, $M=50$, and $R=500$~m: (a) receiver comparison with the dashed $50$--$56$~dB region enlarged; (b) SIC convergence. Markers were digitized from the supplied plots, whose floor is $10^{-4}$.}
		\label{fig:archived_baselines}
		\vspace{-4mm}
	\end{figure*}
	
	\subsection{Successive Interference Cancellation and Channel Re-Estimation}
	\label{subsec:sic}
	
	For verified set $\mathcal K_{\mathrm{succ}}=\{k_1,\ldots,k_S\}$, decoded bits are remapped to $\hat s_{k,j}\in\{+1,-1\}$ and $\mathbf X_k=\mathbf a_k\otimes[1,\hat s_{k,1},\ldots,\hat s_{k,n_c}]$. Stack $\operatorname{vec}(\mathbf X_{k_s})$ as $\boldsymbol\Phi_{\mathrm{pass}}\in\mathbb C^{n_p(1+n_c)\times S}$ and unfold the selected-port residuals as $\overline{\mathbf Y}_{\mathrm{resid}}^{\mathrm{sel}}\in\mathbb C^{n_p(1+n_c)\times M}$. The full-packet RLS estimate is
	\begin{equation}
		\label{eq:cf_h_refine}
		\widehat{\overline{\mathbf{H}}}_{\mathrm{refine}}=
		(\boldsymbol{\Phi}_{\mathrm{pass}}^{\mathrm{H}}
		\boldsymbol{\Phi}_{\mathrm{pass}}
		+\xi\mathbf{I})^{-1}
		\boldsymbol{\Phi}_{\mathrm{pass}}^{\mathrm{H}}
		\overline{\mathbf{Y}}_{\mathrm{resid}}^{\mathrm{sel}},
	\end{equation}
	where $\xi>0$ regularizes noise and residual interference. Unlike a pilot-only estimate, \eqnref{eq:cf_h_refine} exploits the decoded payload while retaining one fixed physical port. Decoded pilots are canceled at all sounded ports, using the refined coefficient at the selected port and the pilot estimate elsewhere:
	\begin{equation}
		\widetilde h_{k,m,q}=\begin{cases}
			\hat{\bar h}_{k,m,\mathrm{refine}},
			& q=q_m^\star,\\
			\hat h_{k,m,q}, & q\neq q_m^\star.
		\end{cases}
		\label{eq:cf_sic_channel_choice}
	\end{equation}
	The pilot and payload residuals are updated as
	\begin{align}
		\label{eq:cf_pilot_residual_update}
		\mathbf{y}_{m,q,0}^{\mathrm{resid}}\leftarrow
		\mathbf{y}_{m,q,0}^{\mathrm{resid}}-
		\sum_{k\in\mathcal{K}_{\mathrm{succ}}}
		\mathbf{a}_k\widetilde h_{k,m,q},\\
		\label{eq:cf_data_residual_update}
		\mathbf{Y}_{\mathrm{resid}}^{(j)}\leftarrow
		\mathbf{Y}_{\mathrm{resid}}^{(j)}-
		\sum_{k\in\mathcal{K}_{\mathrm{succ}}}
		\mathbf{a}_k\hat s_{k,j}
		\hat{\overline{\mathbf{h}}}_{k,\mathrm{refine}}^{\mathrm T}.
	\end{align}
	EM-AMP then processes the pilot residuals again at the fixed port indices. SIC stops when no new packet passes verification or the round limit is reached.
	
	\begin{figure}[!t]
		\centering
		\includegraphics[width=0.8\columnwidth]{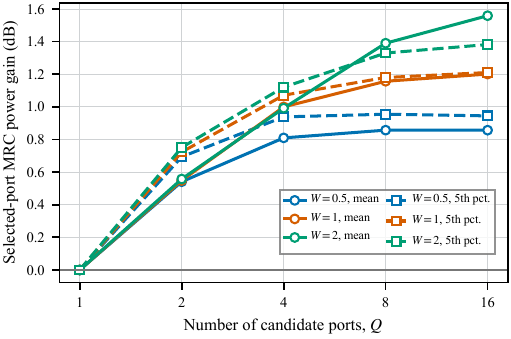}
		\caption{Oracle common-port MRC-power gain over $Q=1$ under perfect CSI. Solid and dashed curves show mean and fifth-percentile gain.}
		\label{fig:fas_gain}
	\end{figure}
	
	\section{NUMERICAL RESULTS}
	
	For clarity, the legend labels in \figref{fig:archived_baselines} and \figref{fig:fas_gain} are defined here. In \figref{fig:archived_baselines}(a), ``Cell-free, SIC + re-est.'' is the complete distributed receiver with SIC and full-packet RLS channel re-estimation; ``Cell-free, SIC only'' performs SIC with the initial pilot-based channel estimates; ``Cell-free, one-shot'' performs a single decoding pass without SIC or channel re-estimation; and ``Colocated, SIC + re-est.'' retains the complete processing chain but colocates all $M$ receive chains at the service center. In \figref{fig:archived_baselines}(b), each curve label specifies $E_b/N_0$, and the horizontal axis gives the SIC round. In \figref{fig:fas_gain}, color identifies the normalized aperture length $W$, solid-circle curves report the mean oracle gain, and dashed-square curves report the fifth-percentile oracle gain relative to the fixed $Q=1$ port.
	
	\paragraph{Simulation Setup and Evaluation Scope}
	
	\tabref{tab:simulation_parameters} lists the common parameters. APs and active UEs are uniform over the permitted disk, with guard-distance violations rejected. Pilot indices are sampled without replacement, spreading is unit-energy, and the centralized baseline colocates the same $M$ independently faded receive chains.
	
	\begin{table}[t]
		\caption{Simulation Parameters}
		\label{tab:simulation_parameters}
		\centering
		\fontsize{6.7pt}{7.3pt}\selectfont
		\setlength{\tabcolsep}{3.2pt}
		\renewcommand{\arraystretch}{1.04}
		\begin{tabular}{@{}p{0.48\columnwidth}p{0.43\columnwidth}@{}}
			\hline
			\textbf{Parameter} & \textbf{Value} \\
			\hline
			Message and pilot index & $B_{\mathrm{info}}=72$, $L_{\mathrm{crc}}=16$; $B_p=12$, $K=4096$ \\
			Coding and payload & 5G NR-LDPC, $n_c=264$, BPSK \\
			Spreading, active UEs, APs & $n_p=150$, $K_a=100$, $M=50$ \\
			Topology and path loss & $R_{\min}=100$ m, $R=500$ m; $d_0=1$ m, $d_{\mathrm{ref}}=20$ m, $\beta=2$ \\
			Outer and SIC iteration limits & 20; 6 \\
			FAS ports and aperture & $Q\in\{1,2,4,8,16\}$, $W\in\{0.5,1,2\}$; nominal $(Q,W)=(8,2)$ \\
			\hline
		\end{tabular}
	\end{table}
	
	\figref{fig:archived_baselines} digitizes the supplied fixed-port plots and is only an archival $Q=1$ reference. \figref{fig:fas_gain} independently evaluates \eqnref{eq:fas_channel_vector}--\eqnref{eq:fas_correlation} using 500 topology/channel realizations and $50{,}000$ normalized power samples per $(Q,W)$. To isolate port selection, perfect-CSI oracle scoring and normalized powers are
	\begin{align}
		&q_m^\star=\arg\max_q\sum_{u=1}^{K_a}|h_{u,m,q}|^2,
		\label{eq:oracle_port_score}\\
		&\Gamma_u^{\mathrm{sel}}=\frac{\sum_m|h_{u,m,q_m^\star}|^2}
		{\sum_m\phi_{u,m}},~
		\Gamma_u^{\mathrm{fix}}=\frac{\sum_m|h_{u,m,1}|^2}
		{\sum_m\phi_{u,m}}.
		\label{eq:fas_normalized_power}
	\end{align}
	The reported gains are
	\begin{align}
		\Delta_{\mathrm{mean}}&=10\log_{10}
		\frac{\mathbb E[\Gamma^{\mathrm{sel}}]}
		{\mathbb E[\Gamma^{\mathrm{fix}}]},
		\label{eq:fas_mean_gain}\\
		\Delta_{0.05}&=10\log_{10}
		\frac{F_{\mathrm{sel}}^{-1}(0.05)}
		{F_{\mathrm{fix}}^{-1}(0.05)},
		\label{eq:fas_gain_metrics}
	\end{align}
	where $F^{-1}$ is the pooled empirical quantile. This oracle benchmark invokes neither EM-AMP nor RLS and is not an end-to-end PUPE result.
	
	\paragraph{Receiver and Port-Selection Results}
	
	Unless otherwise stated, both benchmarks use $K_a=100$ active user-equipment terminals (UEs), $M=50$ single-RF-chain APs (or colocated receive chains), service radius $R=500$~m, UE inner radius $R_{\min}=100$~m, and minimum UE--AP guard distance $d_{\mathrm{ref}}=20$~m. The path-loss normalization distance and exponent are $d_0=1$~m and $\beta=2$, respectively. The fixed-port benchmark sets $Q=1$. PUPE is the per-user probability of error in \eqnref{eq:cf_pupe_metrics}, and $E_b/N_0$ is the energy-per-source-bit to noise-spectral-density ratio in decibels. One SIC round comprises decoding, packet verification, cancellation, and residual re-detection; $10^{-4}$ is the resolution floor of the supplied PUPE plot. For the FAS benchmark, $Q$ is the number of candidate ports per AP and $W$ is the aperture length normalized by the carrier wavelength $\lambda_c$, so the physical aperture is $W\lambda_c$; $(Q,W)=(8,2)$ is the nominal configuration. Perfect channel state information (CSI) is assumed, and selected-port powers are aggregated by maximum-ratio combining (MRC). In \eqnref{eq:fas_mean_gain} and \eqnref{eq:fas_gain_metrics}, $\Delta_{\mathrm{mean}}$ is the dB ratio of the selected-port and fixed-port mean normalized powers, whereas $\Delta_{0.05}$ is the dB ratio of their pooled fifth-percentile normalized powers and characterizes lower-tail reliability. The reported 95\% confidence interval for $\Delta_{\mathrm{mean}}$ is obtained by resampling complete topology/channel-realization blocks.
	
	At 62~dB, \figref{fig:archived_baselines}(a) gives PUPEs of $3.0\times10^{-2}$ for the complete receiver, $6.5\times10^{-2}$ without re-estimation, $7.0\times10^{-2}$ without SIC, and $2.8\times10^{-1}$ for colocated reception. At 64~dB, the respective values are $1.0\times10^{-2}$, $1.8\times10^{-2}$, $3.0\times10^{-2}$, and $1.2\times10^{-1}$. Macrodiversity dominates, while re-estimation improves cancellation. \figref{fig:archived_baselines}(b) shows that most SIC gain occurs within four rounds: at 62~dB, PUPE falls from $7.5\times10^{-2}$ to $3.0\times10^{-2}$, whereas 68~dB reaches the $10^{-4}$ plotting floor after two rounds. Six rounds provide a conservative cap.
	
	\figref{fig:fas_gain} shows that aperture is more valuable than dense sampling of a fixed aperture. For $Q=4$, 8, and 16, $\Delta_{\mathrm{mean}}$ is $0.810$, $0.857$, and $0.857$~dB at $W=0.5$, compared with $0.992$, $1.390$, and $1.559$~dB at $W=2$. At nominal $(Q,W)=(8,2)$, $\Delta_{0.05}=1.330$~dB and the 95\% interval for $\Delta_{\mathrm{mean}}$ is $[1.374,1.408]$~dB. Saturation reflects port correlation; these oracle gains exclude sounding overhead and receiver errors.
	
	\section{CONCLUSION}
	The proposed FAS-assisted cell-free URA receiver combines EM-AMP, port selection, MIMO-IGA, RLS refinement, and SIC. Existing evidence confirms reception, refinement, cancellation, and selection gains, while full FAS PUPE evaluation remains open due to collision, latency, overhead, and scoring challenges.
	\balance

\end{document}